\documentclass[twocolumn]{aastex62}

\usepackage[colorinlistoftodos]{todonotes}
\usepackage{verbatim}
\usepackage[normalem]{ulem}
\usepackage{graphicx}

\graphicspath{{./}{figures/}}

\newcommand{\taurex}{TauREx}
\newcommand{\pylc}{\mbox{PyLightcurve}}

\newcommand{\wfc}{\mbox{HST/WFC3}}

\newcommand{\rprs}{\mbox{R$_p$/R$_\ast$}}
\newcommand{\sma}{a/R$_\ast$}

\accepted{July 28, 2020}
\usepackage{multirow}

\begin{document}
	
	\title{Integrating light curve and atmospheric modelling of transiting exoplanets}
	
	\correspondingauthor{K.H. Yip}
	\email{kai.yip.13@ucl.ac.uk}
	\author[0000-0002-0786-7307]{K.H. Yip}
	\affil{Department of Physics and Astronomy \\
		University College London \\
		Gower Street,WC1E 6BT London, United Kingdom}
	\author[0000-0002-0786-7307]{A. Tsiaras}
	\affil{Department of Physics and Astronomy \\
		University College London \\
		Gower Street,WC1E 6BT London, United Kingdom}
	\author[0000-0002-0786-7307]{I.P. Waldmann}
	\affil{Department of Physics and Astronomy \\
		University College London \\
		Gower Street,WC1E 6BT London, United Kingdom}
	\author[0000-0002-0786-7307]{G. Tinetti}
	\affil{Department of Physics and Astronomy \\
		University College London \\
		Gower Street,WC1E 6BT London, United Kingdom}
	
	\begin{abstract}
    Spectral retrieval techniques are currently our best tool to interpret the observed exoplanet atmospheric data. Said techniques retrieve the optimal atmospheric components and parameters by identifying the best fit to an observed transmission/emission spectrum. Over the past decade, our understanding of remote worlds in our galaxy has flourished thanks to the use of increasingly sophisticated spectral retrieval techniques and the collective effort of the community working on exoplanet atmospheric models. A new generation of instruments in space and from the ground is expected to deliver higher quality data in the next decade, it is therefore paramount to upgrade current models and improve their reliability, their completeness and the numerical speed with which they can be run. In this paper, we address the issue of  reliability of the results provided by retrieval models in the presence of systematics of unknown origin. More specifically, we demonstrate that if we fit directly individual light curves at different wavelengths (L\,-\,retrieval), instead of fitting transit or eclipse depths, as it is currently done (S\,-\,retrieval), the said methodology is more sensitive against astrophysical and instrumental noise. This new approach is tested, in particular, when discrepant simulated observations from HST/WFC3 and Spitzer/IRAC are combined. We find that while S-retrievals converge to an incorrect solution  without any warning, L\,-\,retrievals are able to flag potential discrepancies between the data-sets.
	\end{abstract}
	
	\section{Introduction}
	The characterization of exoplanetary atmospheres is at the forefront of exoplanetary science. 
	The chemical composition and thermal structure of the known exo-atmospheres provide powerful diagnostics to study formation and evolution processes for different classes of exoplanets and, in principle, help to identify habitable worlds.
	
    Until now, the community has relied mostly on instruments in space and from the ground which were not conceived to observe exoplanetary atmospheres, using instruments onboard the Spitzer and the Hubble Space Telescopes or mounted on ground-based facilities such as the VLT, Gemini and Keck observatories. Despite the difficulty in recording minute flux changes, there have been numerous publications reporting the detection of various chemical species in the atmosphere, which include water-vapour, carbon-bearing molecules, alkali metals and ions, condensates and hazes \citep[e.g.][]{charbonneau2002,vidal2004,     Barman2007, tinetti2007, redfield2008, swain2009, Linsky2010, Fossati2010, dekok2013,Barman2015,Macintosh2015,Jacob2018}. When the Wide Field Camera 3 (WFC3) was installed on HST, more and better space-recorded spectra became available \citep[e.g.][]{deming2013,swain2013,stevenson2014,haynes2015,Barman2015}. 
    Despite its narrow spectral range and non-continuous observation window due to the low Earth orbit of HST, molecular species in the atmosphere of exoplanets, such as H$_2$O \citep{berta2012,mandell2013,Ehrenreich2014}, VO/TiO \citep{Evans2016}, NH$_3$ \citep{Macdonald2017} as well as  He \citep{Spake2018} have been identified.
    Years of observations with \wfc \enspace have yielded the first population studies of gaseous planets \citep{sing2016,population} and the very first spectra of super-Earth atmospheres \citep{kreidberg2014,Tsiaras2016,k2-18b}.
    
   To broaden the spectral range covered by existing observations, data from different instruments are often combined \citep[e.g.][]{swain2009, kreidberg2014}. For instance, the IRAC camera on board the Spitzer Space Telescope may offer additional constraints to quantify CO and CO$_2$ abundances and sound the temperature profile of the planet, due to its ability to record mid-IR spectral channels \citep[e.g.][]{cowan2012,zellem2014}. However, current instruments are not calibrated at the 100\,ppm level, and therefore when combining data taken from different observatories, with no overlap in wavelengths, there is the risk of injecting incorrect information in the retrieval.
    An additional issue is that sometimes different data reduction approaches  lead  to different transit/eclipse depths and therefore diverging conclusions in the interpretation, e.g. the case of the thermal inversion in HD\,209458\,b \citep{diamond2014,line2016}. 
    
    Many of the next-generation space missions and ground-based observatories, such as JWST \citep{Greene2016}, ELT \citep{ELT}, TMT \citep{TMT}, Twinkle \citep{edwards_exo} and dedicated missions to exoplanet characterization such as ARIEL \citep{ARIEL} and WFIRST CGI \citep{WFIRST} are due to launch or see first light in the next decade. These instruments promise to achieve a more comprehensive wavelength coverage and/or higher spectral resolving power at a greater precision. The technological advance in instrumentation also prompts the need to upgrade the data analysis techniques to account for uncertainties propagated through the spectral extraction process to the atmospheric model.
    
    Current atmospheric retrieval models have focused on inferring a planet's atmospheric composition and structure by fitting a theoretical spectrum  to an observed transmission/emission/reflection spectrum \citep[e.g.][]{irwin2008,Madhusudhan,chimera,taurex,cubillos2016,helios,ATMO,hydra,al-refaie_taurex3}. This approach does not account for the existing correlation between orbital parameters  and atmospheric components, as only transit/eclipse depths are processed by current retrievals. In other words, the fact that only transit depths are used creates an interruption in the error propagation pathway as transit depths are extracted from the light curves, without accounting for potential correlations between atmospheric transit, and systematics parameters. The complete light curve fitting should be included in the likelihood of the atmospheric retrieval to accurately propagate uncertainties in parameters such as inclination, semi-major axis or limb-darkening coefficients.
    
    In this paper, we focus on transiting planets and propose a novel, more comprehensive approach that takes the retrieval process one step closer to the raw data, by integrating the light curve fitting process into the atmospheric retrieval process. The new approach can uncover systematic errors that were difficult to detect, by correctly accounting for the covariance between atmospheric model and lightcurve fitting.  This paper is  divided into two parts: the first part aims to explain the concept of integrating light curve fitting routines into an atmospheric retrieval framework, followed by case studies to demonstrate the concept in practice and  validate the results. The second part focuses on demonstrating the advantages of using our integrated over the conventional approach, when combining data from different instruments. We investigate the possible effects of systematic errors on the retrieval results. 
    
    \begin{figure*}
		\centering
		\includegraphics[width=0.97\textwidth]{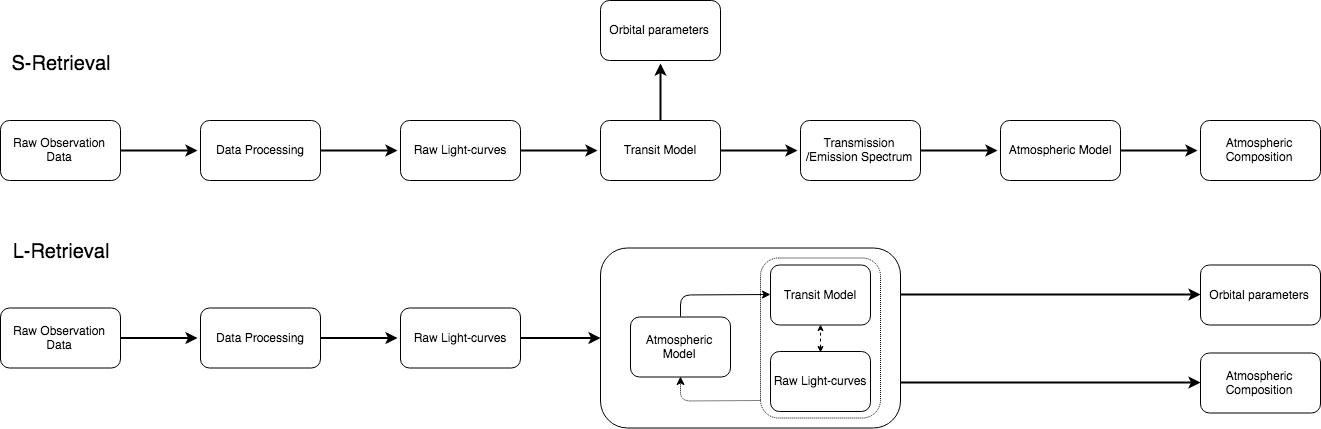}
		\caption{Comparison between the S\,-\,retrieval and our proposed approach, the L\,-\,retrieval. The major difference between the two is the merge of two separate fitting processes into one, bypassing the creation of a transmission/emission spectrum during the fitting process. }
		\label{fig:comparaison}
	\end{figure*}
    
    \section{The L\,-\,retrieval}
    The conventional way (or S-Retrieval) to infer atmospheric components relies on fitting an one dimensional theoretical transmission, emission or reflection spectrum to the observed spectrum. In the case of S-Retrievals the observed spectrum is constructed by extracting the transit depths, and their respective uncertainties, from a series of wavelength-dependent light curves. The resulting one dimensional spectrum is then fitted with the theoretical forward model. This two-step process breaks the error propagation from the light curve fitting to the spectral fitting -- i.e. retrieval -- and the covariances between orbital elements, such as inclination and \sma\ are not taken into account during the spectral retrieval stage.
    
   For clarity, the former approach will be referred to as the spectral retrieval, or `S\,-\,retrieval', hereafter and the latter one as the light curve retrieval, or `L\,-\,retrieval'. See Figure \ref{fig:comparaison} for a schematic comparison between the two approaches.
	
    A light curve records the duration and the extent of the drop in brightness coming from a stellar system, when a planet transits across its host star or it is eclipsed by it. The shape of the light curve is affected by a number of factors, namely, the radius of the planet, the limb-darkening effect from the host star, the orbital configuration of the system and the detectors' characteristics. For the purpose of this paper, which is to illustrate the concept of retrieving the atmospheric parameters directly from light curves, a number of assumptions were made throughout the investigation:
    \begin{itemize}
        \item All the light curves record only a single transit observation.
        \item Observed data is provided in the form of de-trended light curves. The systematics seen in real data are instrument-specific and therefore are omitted in this generalized study.
        \item Building on assumption 1, parameters that cannot be determined from a single transit, such as the period and the eccentricity, are given as constants. 
        \item The limb-darkening coefficients for all light curves are given as constants.
    \end{itemize}
    
    The fitting procedure of the L\,-\,retrieval follows these steps: given a set of raw, de-trended light curves as input, an atmospheric forward model is first generated from a prior distribution of atmospheric components. The simulated binned transit depths are used to generate the set of synthetic light curves. Additional information about the planet-star system is required during this process: e.g. the limb-darkening coefficients, period, eccentricity, inclination, the ratio between the semi-major axis and the radius of the star, \sma \enspace and mid-transit time. As mentioned before, in this investigation we fixed the limb-darkening coefficients, the period and the eccentricity. Once the modelled light curves are generated for different wavelengths, they are compared with the raw light curves to compute the likelihood.

	\begin{figure}
		\centering
		\includegraphics[width=\columnwidth]{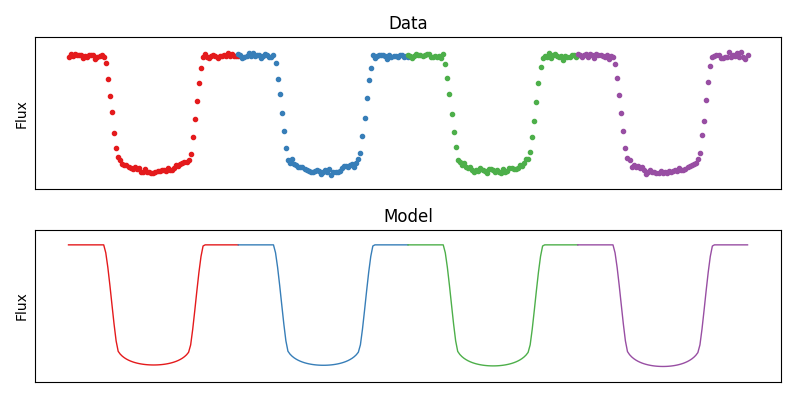}
		\caption{Illustration of the raw light curve chain and the modelled light curve chain. The different colours represent light curves at different wavelength. The x-axis is absent as the light curves are joined together to form an array (chain) during the operation. The actual chain could be of any length and shape depending on the input data.}
		\label{fig:lc_model_cal}
	\end{figure}
	
	To speed-up the fitting process, the simulated light curves at different wavelengths are connected together to form a ``modelled light curve chain". The same process is applied to the raw light curves to form a ``raw light curve chain". A global Gaussian likelihood is computed via equation\,\ref{eqn:lc_chi2}. This equation is empirically, the same as the one used to compute the overall Gaussian likelihood between an observed transit or eclipse spectrum and a modelled spectrum. However, the term $\sigma_{data}$ refers to the uncertainty associated with each data point in the de-trended light curves, and not the one associated with each transit depth. 
	
	{\footnotesize
    \begin{equation}
	    lnL = ln \left [(2\pi \sigma_{data}^2)^{(-n/2)}\text{exp}\left (-\frac{1}{2\sigma_{data}^2}\sum_{i=1}^{n}(x_i-\mu)^2 \right ) \right ]
	    \label{eqn:lc_chi2}
	\end{equation}
    }
    The above process is iterated until it converges to a satisfactory result. Here we implemented the \textit{MultiNest Nested Sampling} routine to sample and map the posterior distribution space for the fitted parameters \citep{Skilling2006,Feroz_multinest}. MultiNest efficiently samples high-dimensional likelihood spaces and has been used extensively by the retrieval community in recent years \citep[e.g.][]{Benneke,Buchner2014,Macdonald2017,helios,ATMO,hydra}. 
    
    The L\,-\,retrieval can be set to either fit for atmospheric components only or fit also for selected orbital elements. In the former case, all the orbital parameters are provided as constants, while in the latter case, $i$, \sma~ and $t_{mid}$ are set as free parameters. The two cases will be denoted as ``orbital fit disabled" and ``orbital fit enabled", respectively, for the rest of the paper.
    
    The implementation of the L\,-\,retrieval is achieved by integrating \pylc \enspace \citep{tsiaras_plc} into \taurex \enspace \citep{al-refaie_taurex3}. The validation of the new retrieval method is discussed in the next section, where the outputs from both S\,- and L\,-\,retrievals are compared. Once the reproducibility and reliability of the L\,-\,retrieval are verified, our investigation focuses on the correlation between the atmospheric and the orbital components.

	\subsection{Reproducibility test} \label{sec:reprod}

	\begin{table}
		\centering
		\caption{Atmospheric Parameters Used To Create The Forward Model of HD\,209458\,b-like planet.}
		\begin{tabular}{ll}
		    \hline \hline
			Atmospheric Parameters                                    & Values                         \\ \hline
			Radius of the Planet ($R_J$)                               & 1.35                          \\
			Cloud Top Pressure (mbar)                                  & 10                             \\
			Water Abundance (log(H$_2$O))                             & -4                            \\
			Temperature of the atmosphere (K)                          & 1500                        \\ \hline
		\end{tabular}
		\label{tb:atm_param}
	\end{table}
	\begin{figure}
		\includegraphics[width=0.49\columnwidth]{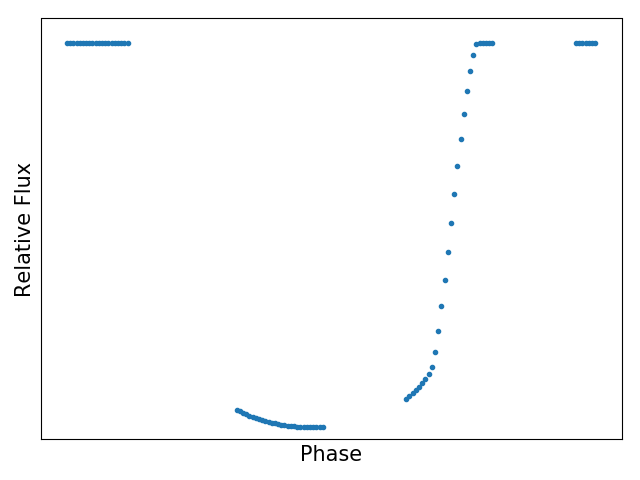}
		\includegraphics[width=0.49\columnwidth]{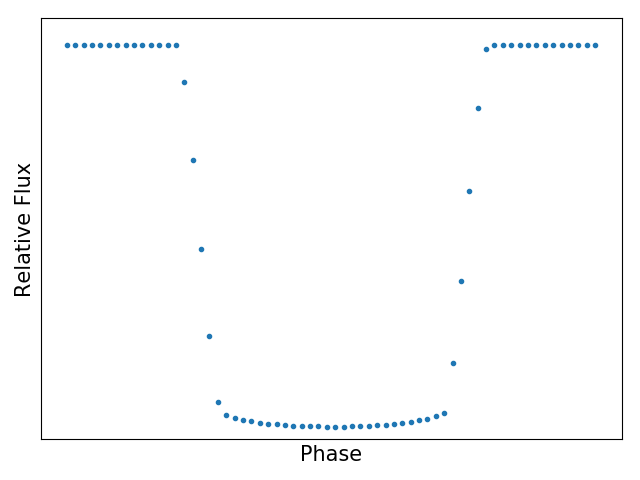}
		\caption{Example of synthetic light curves generated for HST/WFC3 (a) and Spitzer/IRAC (b). Observations taken from HST/WFC3 contained incomplete phase while observations from Spitzer/IRAC captured the entire duration.  The gap in (a) is reproduced according to observation schedule from \cite{deming2013}. 
		\label{fig:lc_3instru}}
	\end{figure}
    
    We simulate the scenario where a transiting event of an HD\,209458\,b like planet is observed via the G141 grism of the HST/WFC3 camera. The input data for this scenario are simulated using both \taurex \enspace and \pylc. Figure\,\ref{fig:lc_3instru} shows a synthetic light curve as recorded with Spitzer/IRAC. The gaps shown in Figure\,\ref{fig:lc_3instru} represent periods of no observations, originated from the low Earth orbit of HST. The timing of the gaps was aligned with data from \cite{deming2013}, to reproduce a more realistic case. 
    A few assumptions were made during the process of generating these light curves:
    \begin{itemize}
        \item The instrumental response is omitted.
        \item The only active gas in the atmosphere is water vapour.
        \item An isothermal Temperature-Pressure profile is assumed.
        \item A grey cloud deck is assumed. 
    \end{itemize}
    Table \ref{tb:atm_param} summarizes the values of the parameters used to generate the forward model. The radius of the planet HD\,209458\,b is taken from \citep{hd209_ref} and the other atmospheric parameters are taken from \cite{Tsiaras2016}. The atmospheric model is binned to match the resolution of the G141 grism. \pylc \enspace is used to generate a set of 25 normalized light curves. To create realistic light curves, the orbital parameters of our simulated system are taken from \citep{hd209_ref}. The limb-darkening coefficients are computed using the PHOENIX stellar models \citep{phoenix,Claret2012,Claret2013}. Finally, a noise level of 100\,ppm is intorduced to the light curves.
	
	\begin{table}
	    \centering
		\caption{Orbital Configuration of HD\,209458\,b-like planet \citep{hd209_ref}.}
		\begin{tabular}{ll}
			\hline \hline
			$P$                       & 3.524                  \\
			$e$                 & 0                      \\
			$i$                  & 86.71                  \\
			\sma          & 8.76                   \\
			T$_{mid}$            & 2456196.28836          \\
			Periastron                   & 0                     \\ \hline
		\end{tabular}
		\label{table:orb_config}
	\end{table}

	The data are fed into both S- and L-retrievals. Both modes of L\,-\,retrieval were tested to assess their validity. 
	
    An additional treatment is needed for the S\,-\,retrieval. The simulated data underwent individual Markov Chain Monte Carlo \citep[MCMC][]{MCMC} fitting to extract the transit depth at each wavelength. The incomplete coverage of the HST/WFC3 data has made it difficult to fit for both \rprs \enspace and orbital elements due to their intrinsic degeneracy. As the orbital configuration of the system is known a priori, the MCMC is set to fit for \rprs \enspace only. The extracted transit depths are used as input to the S\,-\,retrieval.
	
	\subsubsection{Reproducibility Test Result}
	\label{sec:validity}
	\begin{figure}
    \centering
		\includegraphics[width=\columnwidth]{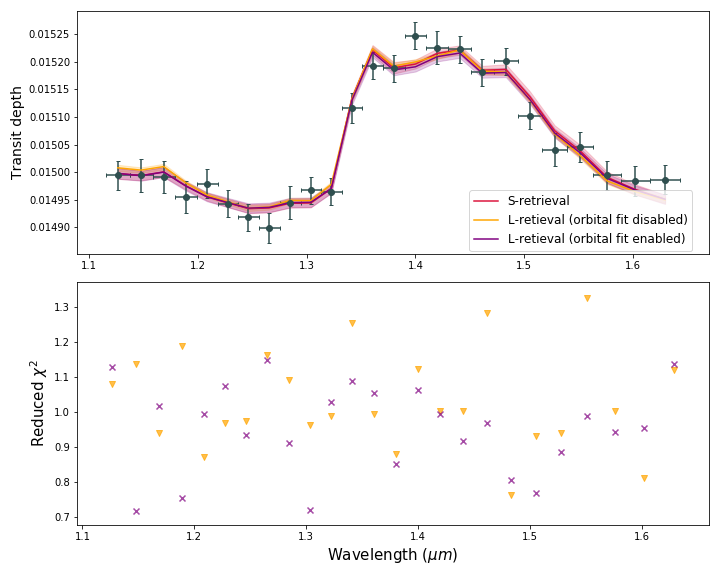}
		\caption{Theoretical transmission spectra retrieved by the two approaches. The red spectrum is retrieved using the S\,-\,retrieval. The orange spectrum is retrieved using the L\,-\,retrieval, with orbital fit disabled. The purple spectrum is also retrieved from the L\,-\,retrieval, but with orbital fit enabled. The three spectra agreed well with each other. The lower plot displays the reduced $\chi^2$ between the raw and model light curves at different wavelengths. This is to highlight any remaining residuals between each pairs of raw and model light curves. The colour of the dots follows the colour code on the upper plot, red dots are omitted as they are retrieved using the S\,-\,retrieval. }
		\label{fig:lc_normal_spectrum_noorbit}
	\end{figure}
	
    Figure \ref{fig:lc_normal_spectrum_noorbit} shows the transmission spectra retrieved by the S\,-\,retrieval (red), and the L\,-\,retrievals with orbital fit enabled (purple) and disabled (orange). The shaded region represents the 1\,$\sigma$ confidence interval. Note that only the light curves are used as an input for the L\,-\,retrievals (purple and orange) and not the transit depths. The observed transit depths were plotted to help the reader evaluate the agreement between the theoretical transmission spectra and the observed transit depths. The lower plot displays the reduced $\chi^2$ between the raw data and the model light curves within each spectral bin, to highlight any remaining residuals. The residual time series for this case and other investigations can be found online\footnote{https://osf.io/kt2e3/}.
    
\begin{figure}
		\centering
		\includegraphics[width=\columnwidth]{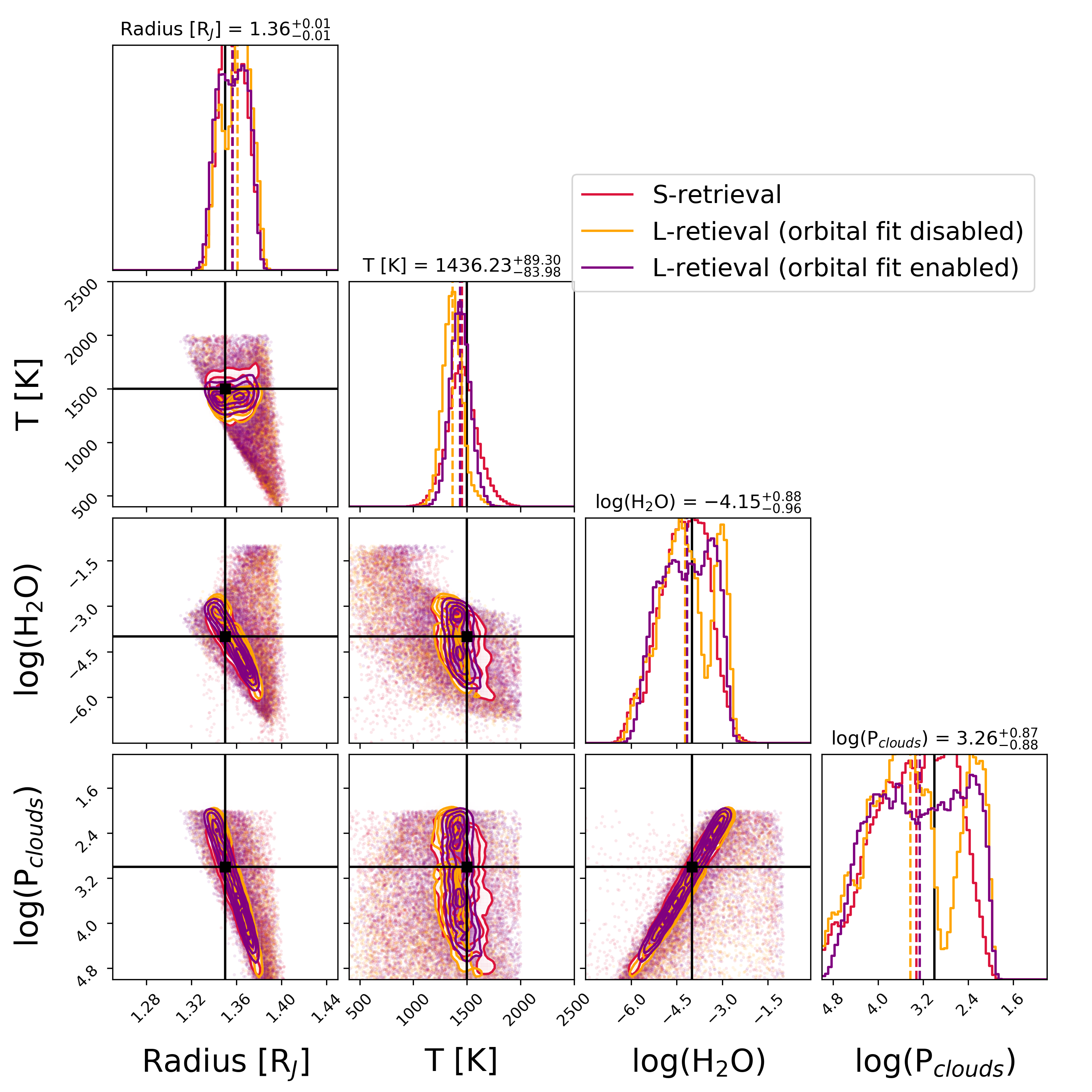}
		\caption{Posterior distributions of all three retrievals in Figure \ref{fig:lc_normal_spectrum_noorbit}. Only the common parameters between the posterior distributions are plotted. The colour code of each contours follows the colour code in Figure \ref{fig:lc_normal_spectrum_noorbit}. The three retrievals obtained consistent result with each other and are in line with the ground truth (indicated by the black solid line). }
		
		\label{fig:lc_normal_posterior_noorbit}
	\end{figure}

    Figure~\ref{fig:lc_normal_posterior_noorbit} shows the corresponding posterior distributions of the different retrieval processes, with the same color code as Figure \ref{fig:lc_normal_spectrum_noorbit}. These posterior distributions (and similar plots thereafter) are all generated as a product of nested sampling. All three contours converged to consistent result with the input parameters (see Table \ref{tb:atm_param}) and with each other. The L\,-\,retrieval, however, showed tighter bounds in various parameters.
    Possible reasons behind the tighter uncertainty bounds are discussed in Section \ref{sec:uncertainty}.
	\subsection{Correlation between the orbital parameters and atmospheric components}
    \label{sec:corr}
	\begin{figure}
		\centering
		\includegraphics[width=\columnwidth]{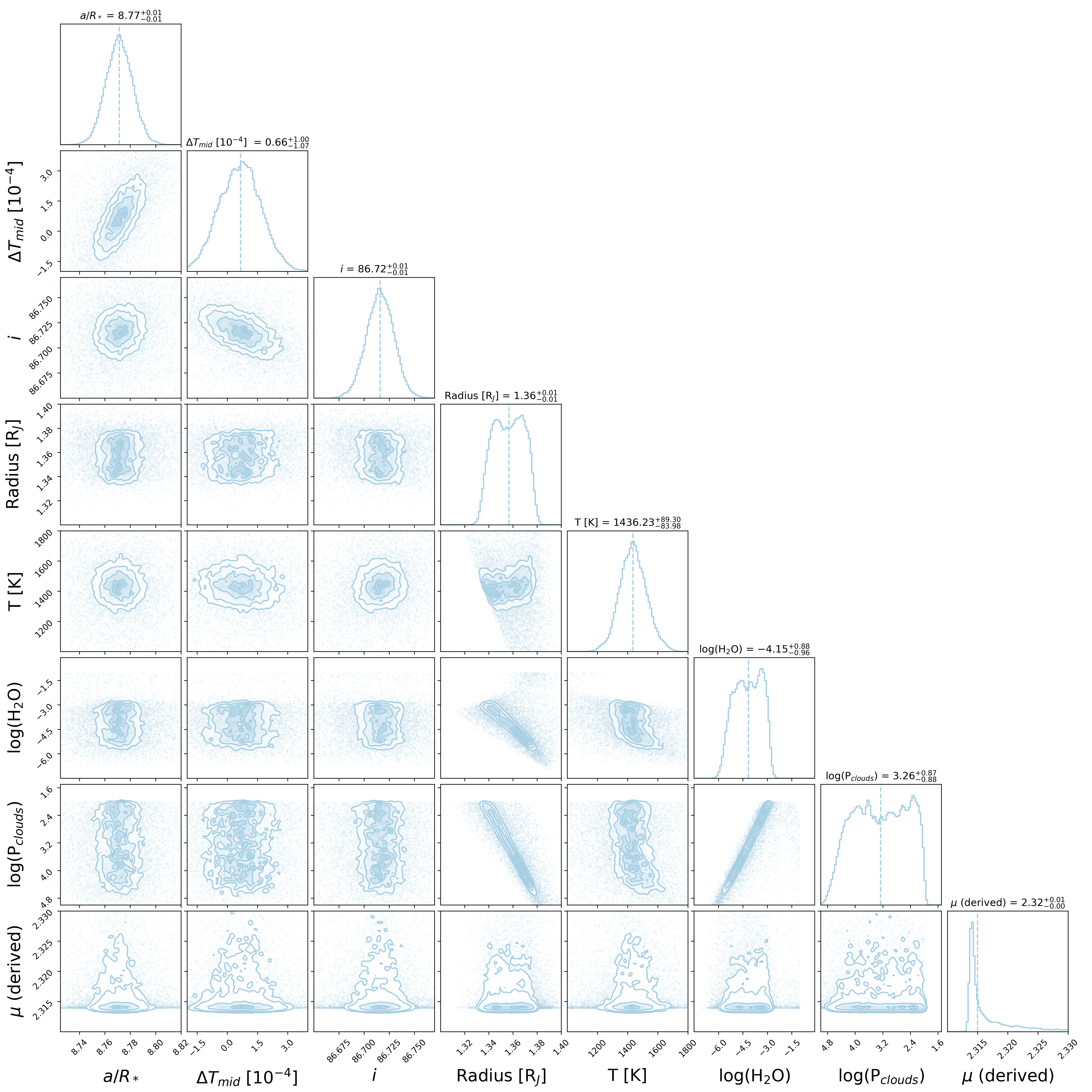}
		\caption{Posterior distribution of the  parameters retrieved using the L\,-\,retrieval, with selected orbital elements , i.e. mid-transit time, inclination and \sma \enspace taken as 
		parameters. The upper three diagonal components represent the orbital elements' contribution and the lower four components represent the atmospheric contribution.  The retrieved T$_{mid,r}$ could be computed by adding T$_{mid}$ from Table \ref{table:orb_config} to $\Delta$T$_{mid}$}
		\label{fig:lc_posterior}.
	\end{figure}
    Figure \ref{fig:lc_posterior} shows the posterior distribution of the L\,-\,retrieval, with orbital fit enabled. The top three diagonal elements show the distribution of the orbital parameters and the lower four show the distribution of the atmospheric elements. We found no significant correlations in the conditional distributions of the atmospheric and orbital parameters. 
	
	\section{Behaviour of L\,- and S\,-\,retrievals in multi-instruments retrievals}
    
    The following case studies look at the possible impacts to both retrieval methods when data from different instruments are combined together during a retrieval. 
    
    We simulated observations of our HD209458b-like planet from two instruments, HST/WFC3 G141 grism (1.1 - 1.8$\mu$m) and Spitzer/IRAC (3.6, 4.5, 5.7 and 7.8\,$\mu$m). 
    
    We followed the same data generation process as described in Section \ref{sec:reprod}. The same atmospheric and orbital parameters ($P$, $e$, $T_{mid}$ and argument of periastron) were used to simulate light curves for HST/WFC3 and Spitzer/IRAC (see Figure \ref{fig:lc_3instru} for a comparison between light curves obtained from HST and Spitzer). Wavelength dependent limb-darkening coefficients were computed using PHOENIX stellar models for HST/WFC3 wavebands as before and the coefficients for the 4 photometric channels from Spitzer/IRAC were taken from \cite{spitzer_limb}.
    
    The simulated observations were then supplied to both L- and the S- retrievals, for comparison. Similar to the previous case, we supplied the correct orbital configuration during the light curves extraction process and only fit for the \rprs \enspace using MCMC for both WFC3 and Spitzer/IRAC spectral range. 
    
    We investigate five possible scenarios when data from both instruments were combined: 
    \begin{enumerate}
    \item Unbiased observation for both instruments.
    \item  Biased observation for HST/WFC3 data (wrong orbital parameters).
    \item  Biased observation for Spitzer/IRAC data  (wrong orbital parameters).
    \item  Biased observation for HST/WFC3 data (wrongly normalised lightcurves with wavelength-dependent ramps).
    \item Wrong model assumption in atmospheric models (ammonia included in simulation but not in the retrieval).
    \end{enumerate}
    
    Below we will describe the individual setup for each scenario and their corresponding outcomes. A summary table of the setup could be found in  Table \ref{table:orbital_config}. 

    \begin{table*}[]
    \tiny
    \caption{Details of the scenarios explored when observation from HST/WFC3 and Spitzer/IRAC were combined. Bold text highlights the change in $i$ and $a/R_*$. Unstated parameters are left intact.}
    \begin{tabular}{cccccccccc}
    \hline
    \multirow{2}{*}{\begin{tabular}[c]{@{}c@{}}Orbital \\ configuration\end{tabular}} & \multicolumn{9}{c}{Scenarios} \\
     & \multicolumn{2}{c}{1} & \multicolumn{2}{c}{2} & \multicolumn{2}{c}{3} & \multicolumn{2}{c}{4} & 5 \\ \hline
    Instruments & HST/WFC3 & Spitzer/IRAC & HST/WFC3 & Spitzer/IRAC & HST/WFC3 & Spitzer/IRAC & HST/WFC3 & Spitzer/IRAC & HST/WFC3 \\
    Active gases & \multicolumn{2}{c}{H$_2$O} & \multicolumn{2}{c}{H$_2$O} & \multicolumn{2}{c}{H$_2$O} & \multicolumn{2}{c}{H$_2$O} & H$_2$O,NH$_3$ \\
    Systematics? & No & No & No & No & No & No & Yes & No & No \\
    $i$ & 86.71 & 86.71 & \textbf{87.21} & 86.71 & 86.71 & \textbf{87.21} & 86.71 & 86.71 & 86.71 \\
    \sma & 8.76 & 8.76 & \textbf{8.86} & 8.76 & 8.76 & \textbf{8.86} & 8.76 & 8.76 & 8.76 \\
    \hline
    \end{tabular}
    \label{table:orbital_config}
    \end{table*}
    
    \paragraph{Scenario\,1}
    This scenario is an extension of the reproducibility test in Section \ref{sec:reprod}, which served as a reference point for other scenarios. Both instruments (HST and Spitzer) are generated with correct orbital configuration.
    
    \paragraph{Scenario\,2}
    Biased Observations are received from HST/WFC3. These observations are generated with wrong orbital parameters. In particular, \sma and $i$ are manually increased by 0.5 and 0.1 respectively.  The offsets was arbitrarily chosen to provide reasonable deviation from the ground truth. Spitzer/IRAC, observation on the other hand, were left intact. 
    
    \paragraph{Scenario\,3}
    Biased Observations are received from Spitzer/IRAC. These observations are simulated using the same procedure outlined in Scenario 2. HST/WFC3 observations were left intact in this case.
    
    \paragraph{Scenario\,4}
    In this scenario we introduce biases due to uncorrected systematics in the HST/WFC3 time series data.
    We simulated inclined light curves, with a wavelength dependence on the slopes’ amplitudes (the shorter the wavelength the steeper the slope), a behavior commonly found with WFC3 data. Additionally, we have added an offset to the normalisation factor, to simulate out-of-transit normalisation errors. 

    \paragraph{Scenario\,5}
    To understand how the L-retrieval will respond to unknown atmospheric opacities, we created an atmosphere with H$_2$O and NH$_3$ as active gases and assumed a H$_2$O only atmosphere during retrievals

    Results of the above scenarios are summaries in Figure~\ref{fig:spectrum_panel}. For each scenario we have included the retrieved spectra coming from S-retrieval, L-retrieval (orbital fit enabled) and L-retrieval (orbital fit disabled) in red, orange and purple respectively. The black dotted line represents the ground truth in each case. The goodness of fit for each case is evaluated via reduced $\chi^2$ calculation between the model and the data (Right column). Specific results of each scenario will be discussed below. 
	
	\subsection{ Scenario\,1: Unbiased observation for both instruments}
    \label{sec:scen1}	
    
	Both retrieval methods are able to retrieve the correct atmospheric parameters within one sigma.  The slight offset from the ground truth is due to the injection of Gaussian noise during the generation process of the light curves.
	Figure\,\ref{fig:pt2_w3_sp_correct_spectrum_poster} shows the corresponding posterior distributions of both types of retrievals. As expected, both retrievals converge to the same solution. 
	
	\subsection{ Scenario\,2 and 3: Biased observation from either instruments}
	\label{sec:scen2}
	
\begin{figure}
		\centering
        \includegraphics[width=\columnwidth]{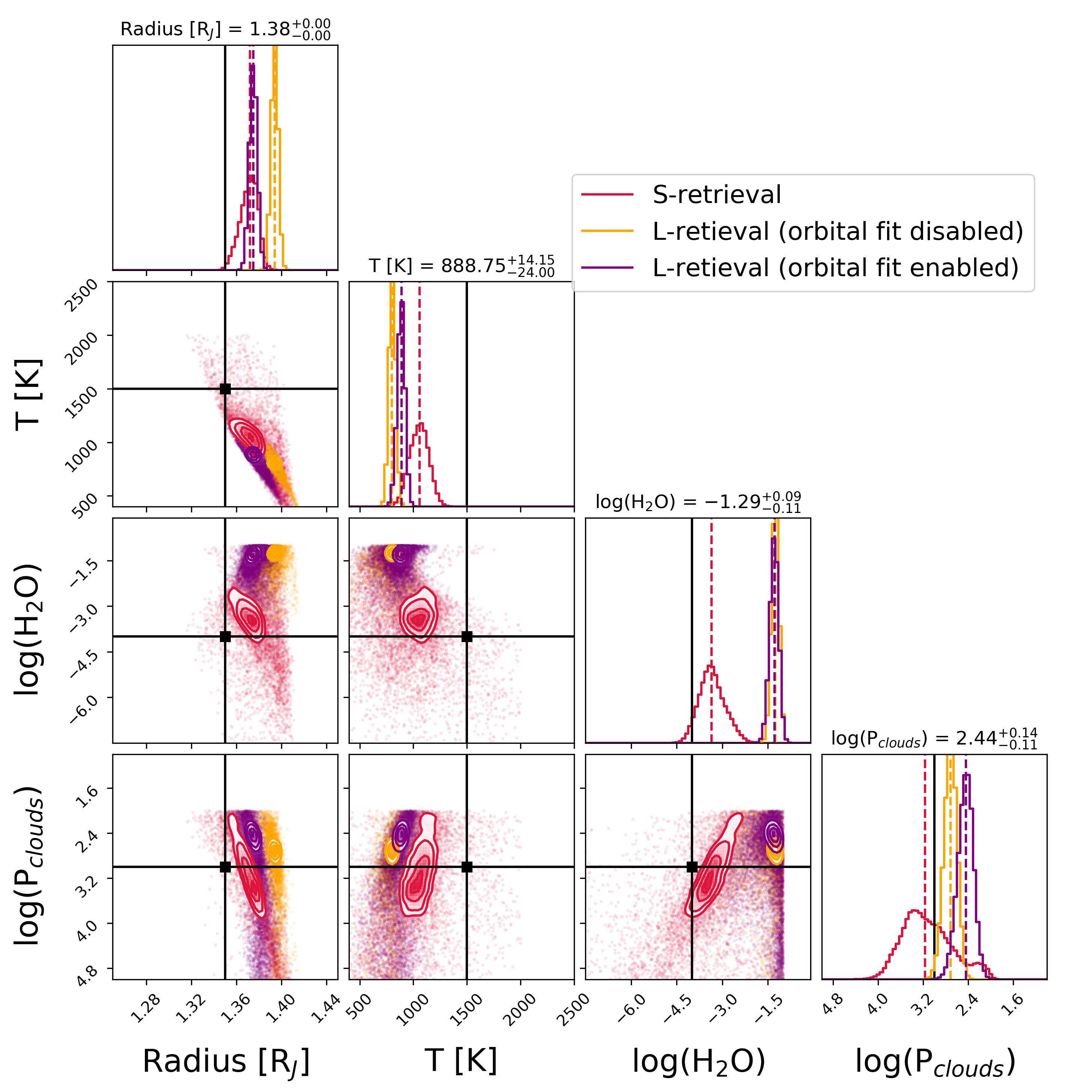}
		\caption{Posteriors distributions of the two retrieval techniques from Scenario 2, with the common retrieved parameters plotted together. All the retrievals did not converge to the ground truth (indicated by the black solid line).  }
		\label{fig:pt2_wrong_w3_sp_pos}
	\end{figure}

\begin{figure}[]
    \centering
     \includegraphics[width=0.9\textwidth]{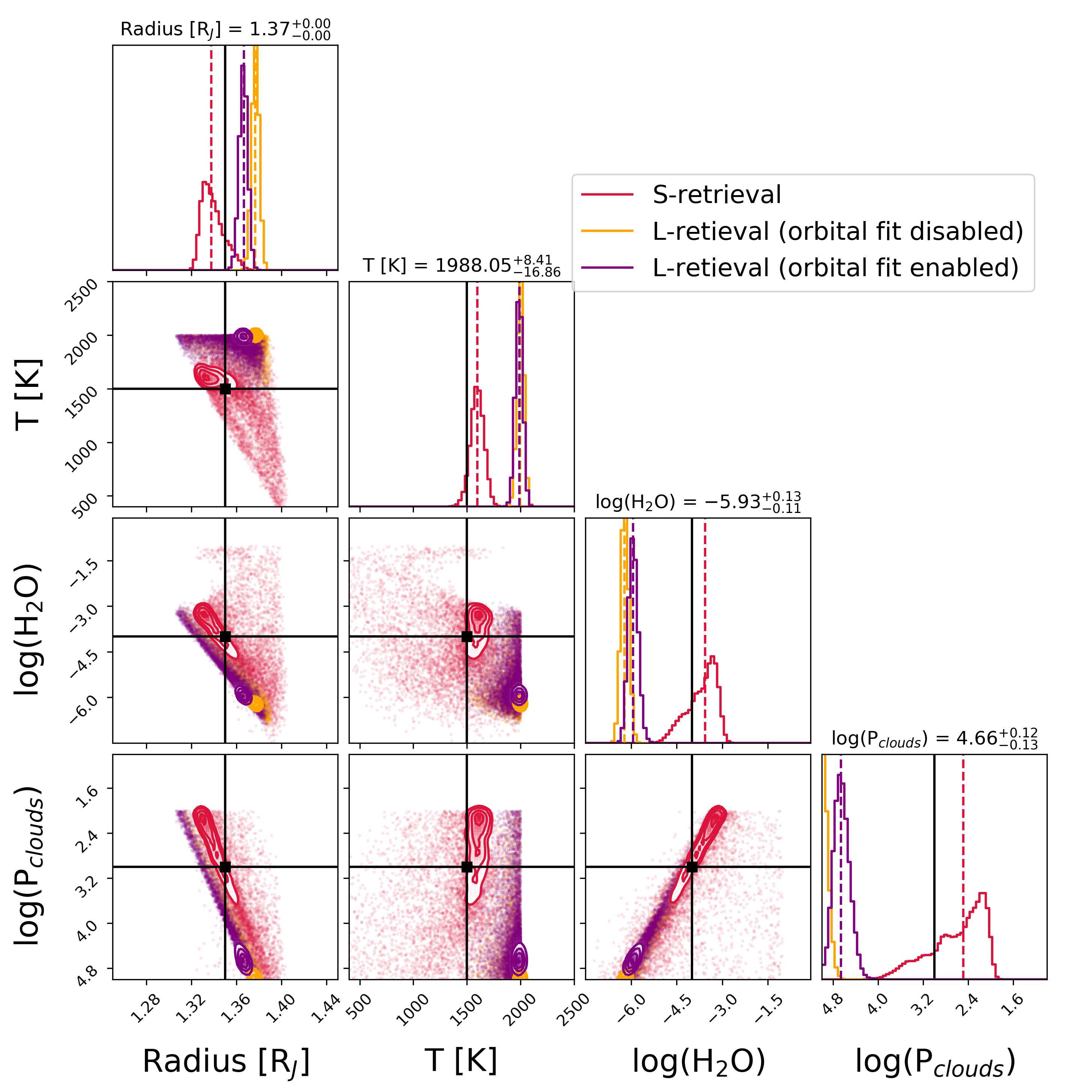}
    \caption{Posteriors distribution of the two retrieval techniques from Scenario 3. The black solid line indicates the ground truth value. Unlike Scenario 2, retrieved values from s-retrieval are within 1-$\sigma$ of the ground truth. This is likely to be due to the overwhelming number of correct HST/WFC3 datapoints in the spectrum.  }
    \label{fig:scen3_pos}
    \end{figure}

	Unlike Scenario\,1, responses from the S\,- and L\,- retrievals diverge in both scenarios. While the S\,-\,retrieval presented the best fit atmospheric components according to the data points, the L-retrieval  did not align with the data points even with or without orbital fit enabled. In Figure~\ref{fig:spectrum_panel}, the reduced-$\chi^2$ is significantly higher than unity for both types of L-retrievals, indicating a poor fit to the data. In the orbital fit enabled L-retrieval, we can see the code trying to fit to the HST data but failing to fit Spitzer data. This is not entirely unexpected as most data is contained in the HST observations. As we can clearly see in the posterior distributions of the parameters retrieved (Figure \ref{fig:pt2_wrong_w3_sp_pos} \& \ref{fig:scen3_pos}), it is noteworthy that none of the retrieved solutions match with the ground-truth forward model. Whilst this is not unexpected either, it showcases that L-retrievals can at least indicate fitting/offset issues between data sets
    
    \subsection{Scenario 4: wrongly normalised lightcurves with wavelength-dependent ramps in HST/WFC3 data}
    
\begin{figure}[]
    \centering
     \includegraphics[width=0.9\textwidth]{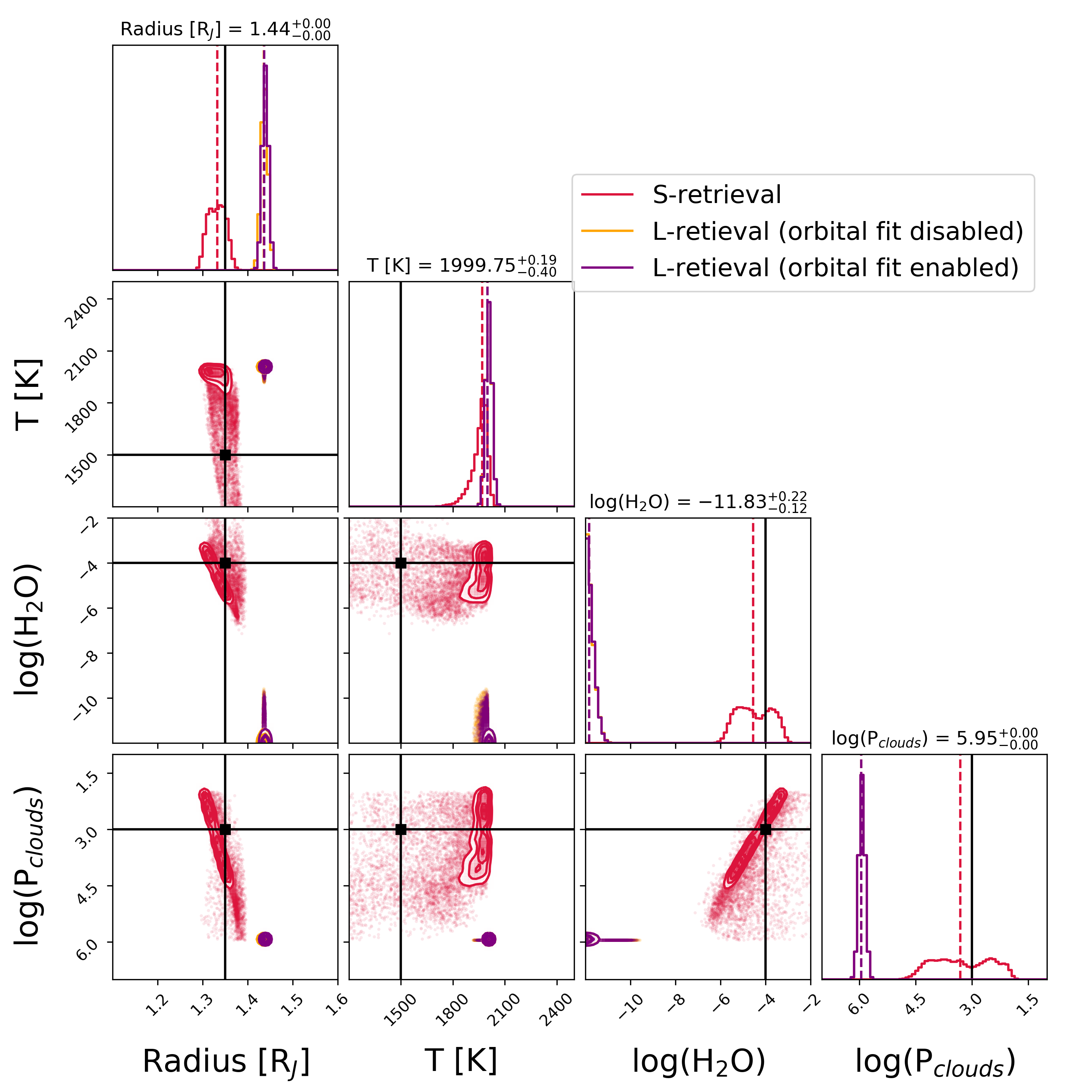}
    \caption{Posteriors distribution of the two retrieval techniques from Scenario 4. The black solid line indicates the ground truth value. Results from L-retrieval are strongly deviated from the ground truth. }
    \label{fig:scen4_pos}
    \end{figure}

    Similar to the outcome from Scenario 2 and 3, neither L- nor S-retrievals are consistent with one another and neither were able to retrieve the correct values, see Figure~\ref{fig:spectrum_panel} (4th row). As before, the S-retrieval erroneously fits most spectral points, resulting in a highly biased result. The L-retrievals returned very high reduced $\chi^2$ values correctly indicating that a consistent fit to the data is not possible (See Figure \ref{fig:scen4_pos} for corresponding posterior distribution). 
    
    
    
    \subsection{ Scenario 5: Wrong model assumption in atmospheric models }
\begin{figure}[t!]
    \centering
     \includegraphics[width=0.9\textwidth]{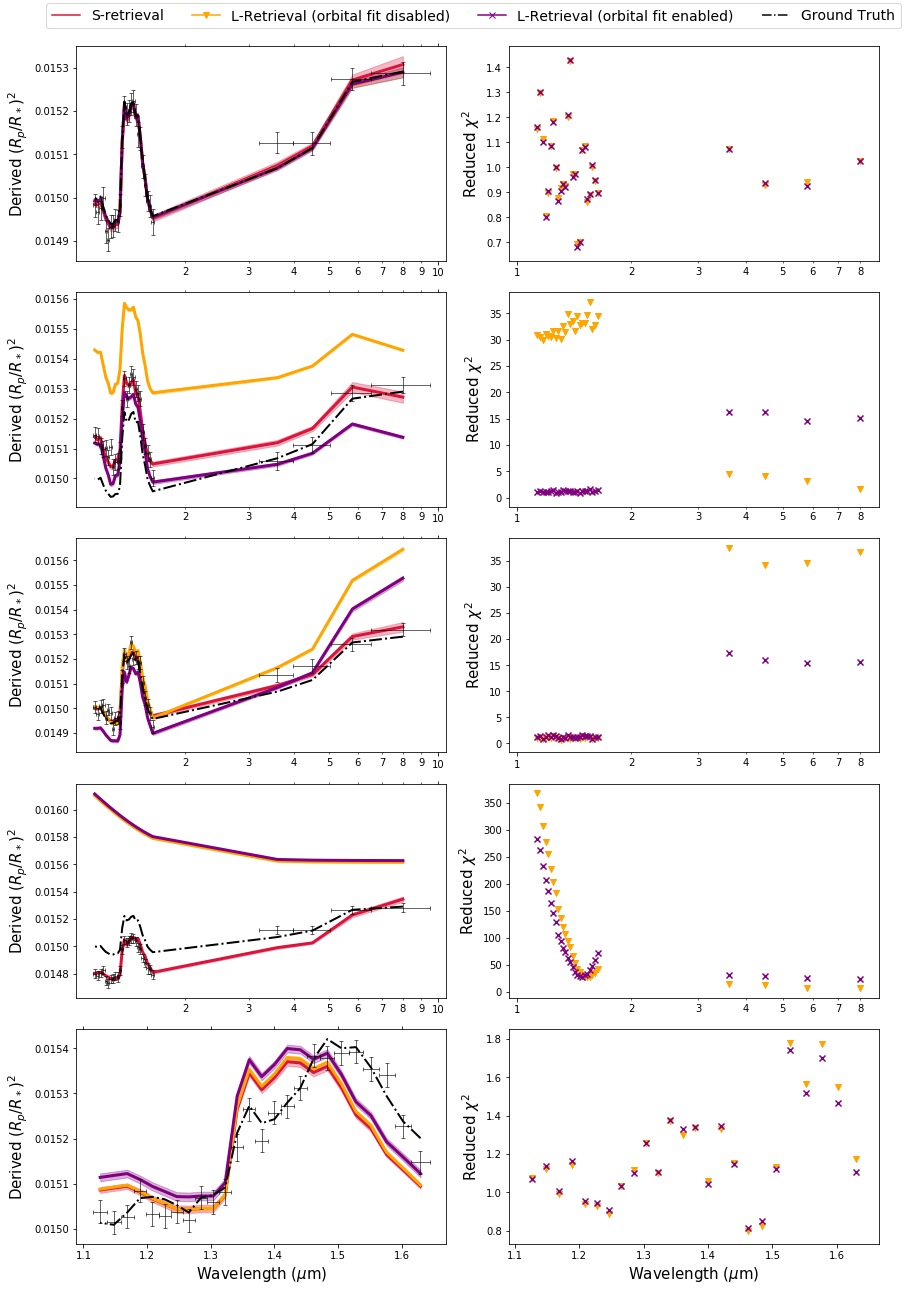}
    \caption{Panel showing the results from all 5 scenarios (Top to bottom: Scenario 1 to Scenario 5) as outlined in Section 3 of the main text. The first column shows the retrieved spectra using different retrieval methods and settings. The second column shows the reduced $\chi^2$ between the raw light curve and the model light curve. Results from S-retrieval, L-retrieval (orbital fit disabled) and L-retrieval (orbital fit enabled) are coloured in red, orange and purple respectively. }
    \label{fig:spectrum_panel}
    \end{figure}

\begin{figure}[]
    \centering
     \includegraphics[width=0.9\textwidth]{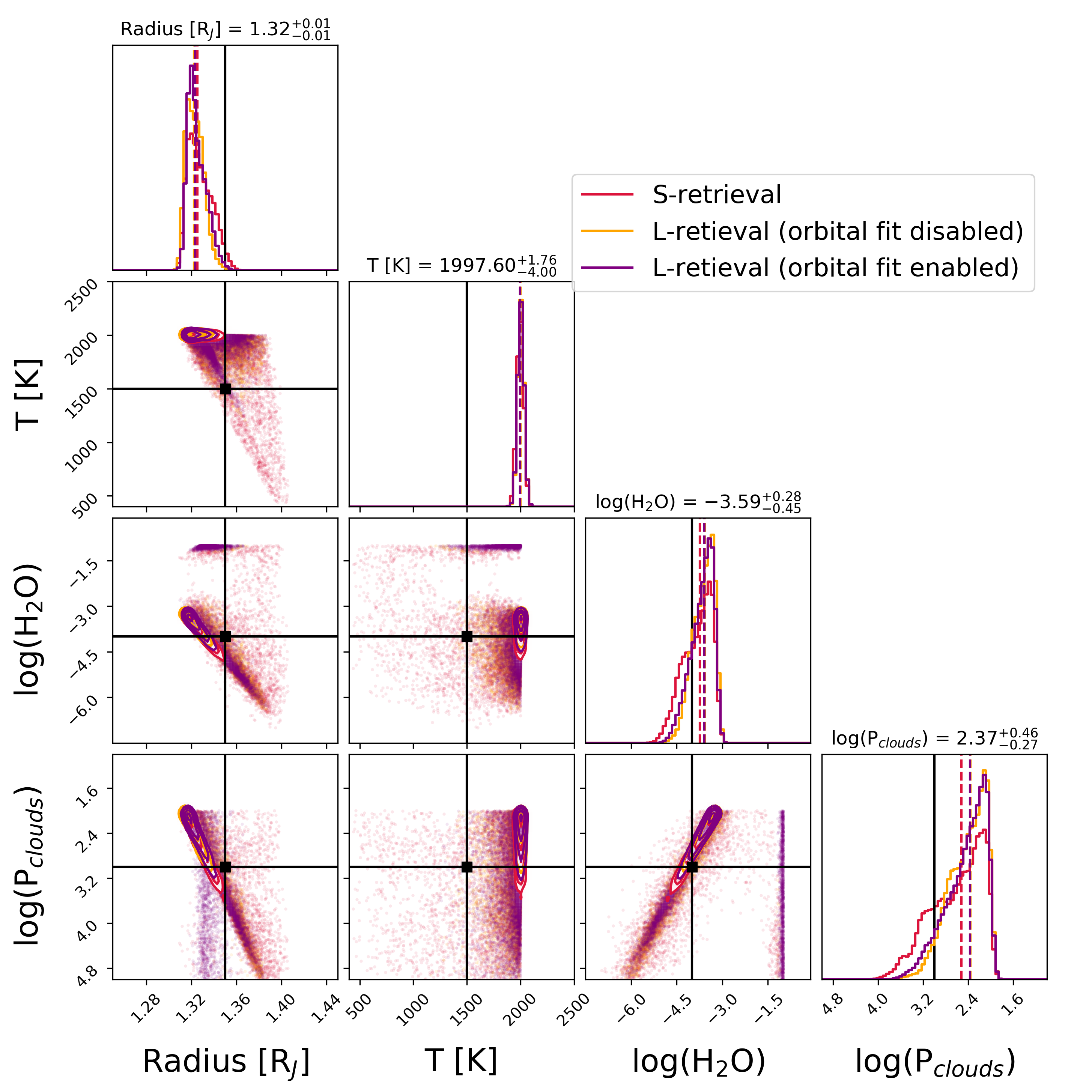}
    \caption{Posteriors distribution of the two retrieval techniques from Scenario 5. The black solid line indicates the ground truth value. Here both retrieval techniques behaved similarly and returned the wrong results. }
    \label{fig:scen5_pos}
    \end{figure}

    Here we test the retrieval's robustness to unseen atmospheric opacities by excluding NH$_3$ from the retrieval and only fitting for H$_2$O, bottom panel in Figure~\ref{fig:spectrum_panel}. Unsurprisingly, both S- and L- retrievals obtain very similar best-fit forward models. It provides evidence that both retrieval techniques reacted similarly when `outliers' are coming from the transit depth (i.e. an incomplete forward model) instead of the shape of the light curve. This finding further reinforces the idea that L-retrieval is an extension of the S-retrieval, with the additional sensitivity to the shape of the light curves(See Figure \ref{fig:scen5_pos} for corresponding posterior distribution).

	\section{Discussion} \label{Discussion}
    \subsection{Uncertainty on retrieved parameters}
    \label{sec:uncertainty}
\begin{figure}
		\centering
	    \includegraphics[width=\columnwidth]{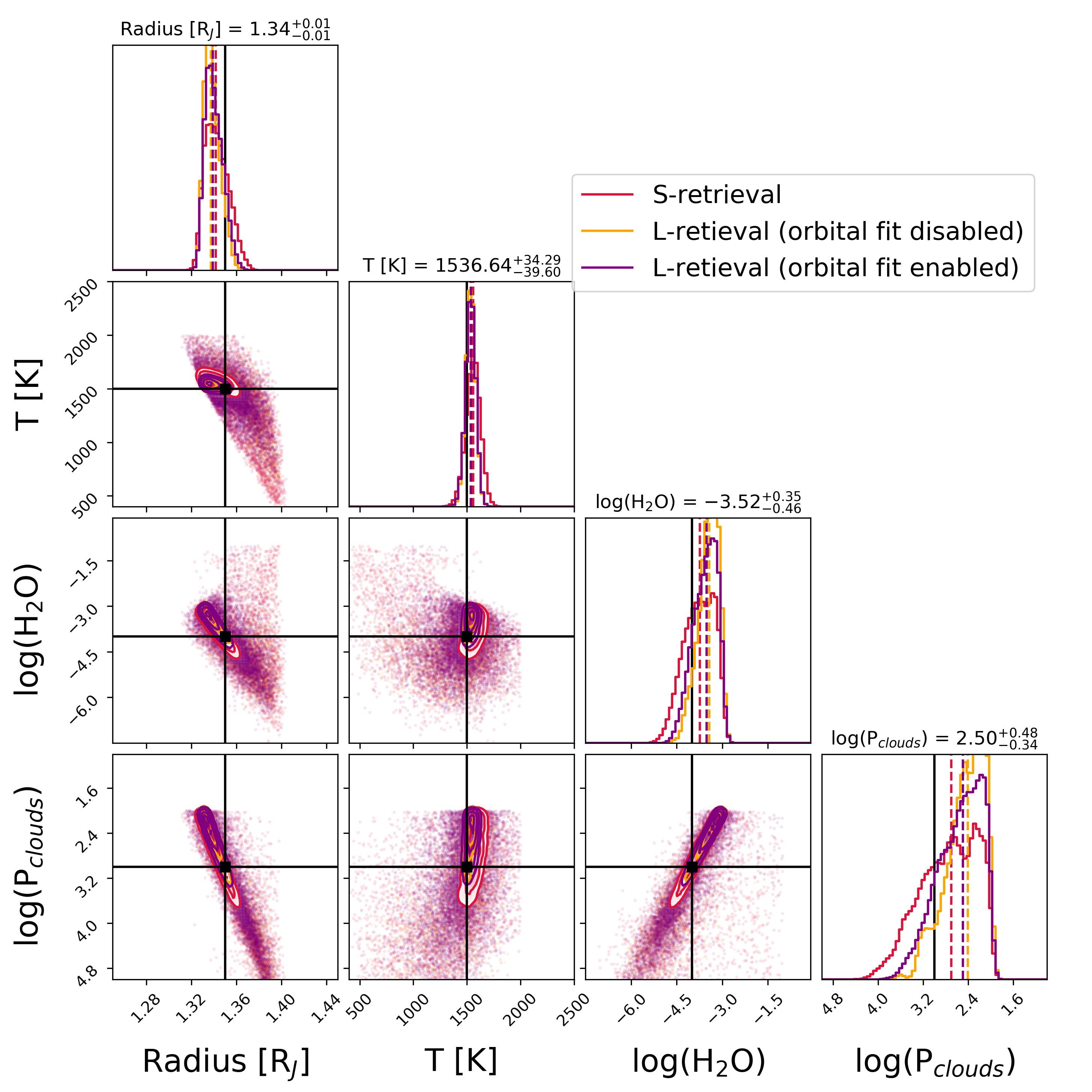}
		\caption{Posteriors distributions of the two retrieval techniques from Scenario 1, with the mutual retrieved parameters plotted together. All three of them were consistent with the ground truth (indicated by the black solid line). }
		\label{fig:pt2_w3_sp_correct_spectrum_poster}
	\end{figure}

    Results obtained from the reproducibility test and Scenario 1 (Figure~\ref{fig:pt2_w3_sp_correct_spectrum_poster}) showed that both L\,- and S\,- retrievals were consistent with each other. The two retrievals, however, possessed different levels of uncertainty, with the L\,-\,retrievals placing a very slightly tighter bound on the retrieved parameters. We argue that the L\,-\,retrieval provides more realistic, consistent estimates on the uncertainties 
    for the following reason: The move to fit directly on the light curves allows more accurate error propagation, as it is one step closer to the observed data. Any systematic errors that remain in the light curves after the data calibration and extraction process are evaluated jointly and reflected in the atmospheric components. Between the two modes in L\,-\,retrieval, the posterior distribution with orbital fit enabled (purple),  features larger uncertainty bounds than posterior distributions with orbital fit disabled (orange). The wider uncertainty bounds are likely due to the increase in the dimensionality of the model.  
    
    \subsection{Oblivious behaviour of S-retrievals in Scenario 2, 3 \& 4 }
    \label{dis:offset}
    
\begin{figure}[]
    \centering
    \includegraphics[width=\columnwidth]{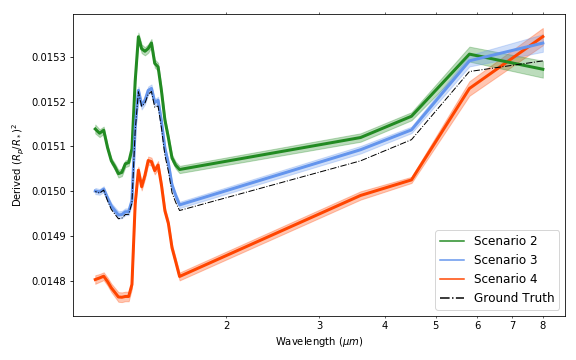}
        \caption{Comparison between the transmission spectra retrieved in Scenario 2, 3 and 4 using the S-retrieval and the ground truth. The forward model is unaltered in all of these cases, however, solutions given by the S-retrieval all failed to retrieve the correct spectrum. }
        \label{fig:pt2_obs_compare}
    \end{figure}

    Throughout most of the scenarios, S-retrieval has closely followed the data points given by the observed spectrum, which is expected from a model fitting perspective. However, we argue that this behaviour is a potential weakness of the S-retrieval. 
    Figure \ref{fig:pt2_obs_compare} compares the results of S-retrieval from Scenarios 2 to 4. None of the results managed to retrieve the correct spectrum.
    This outcome thus starkly demonstrates the perils of classical S\,-\,retrievals when orbital parameters are poorly constrained. For example, light curves with incomplete phase coverage (such as Hubble observations) strongly rely on the input of pre-determined orbital configurations in order to extract \rprs. However, orbital configuration taken from external sources may include biases, in particular in the limb-darkening assumed. Such biases in orbital parameters can lead to significantly differences in the atmospheric retrieval solutions. Since the S\,-\,retrieval is disconnected from the lightcurve fitting, detecting such biases is not possible, whilst it is a natural outcome of the L\,-\,retrieval method. 
    
    \subsection{Behaviour of L-retrieval in Scenario 2,3 and 4 }
    \label{dis:contradict}
    
    Many of our tests have seen L-retrieval not being able to converge to a reasonable fit, compared to its S-retrieval counterpart. Within all of these scenarios we have only altered the geometric information of the lightcurve (orbital elements or slopes of lightcurves) and have left the atmospheric forward model intact. 
    
    As lightcurves are directly fed as input in the case of L-retrieval, it is  forced to deal with these potential bias.
   The geometric information of the light curves, in these cases, has imposed an additional penalty on the retrieval. 
    
    Scenarios 2 \& 3 are a demonstration of the tug of war between Hubble and Spitzer data when fitted as 1D spectra (S-retrieval) or light curves (L-retrieval). As previously discussed, the S-retrieval will settle on a solution closest to the 1D spectrum and may hence be biased by `unseen' and unpropagated systematic errors in the light-curve fitting. On the other hand, the geometric penalty is preventing the L-retrievals from converging. We thus argue that the L-retrieval is a technique that is sensitive to the compatibility of the data, for example in the case of combining different instruments together. It helps to flag potential issues with the combined data and indicate incompatibility of data sets.

    \section{Conclusion}
    In this paper we have introduced a new, integrated approach to retrieve exoplanet atmospheres by integrating light curve fitting into the classical retrieval approach. We have demonstrated that this approach, the here called L\,-\,retrieval,  has significant advantages over the conventional retrieval on an 1D spectrum (S\,-\,retrieval) and is complementary to the S\,-\,retrieval which helps to improve the reliability of the retrieved result. 
    By fitting directly the light curves, we can propagate the systematic uncertainties and parameter correlations from the light curves to the estimate of the atmospheric parameters. We find the L\,-\,retrieval to be more sensitive to correlations in the parameter space and to generally yield tighter parameter constraints compared to the classical S\,-\,retrieval. 
    
    When combining data of multiple instruments or epochs, bias offsets are possible due to systematic errors in the instrument calibration, stellar noise, or poorly constraint orbital parameters. The S\,-\,retrieval is oblivious to the constraints in the orbital parameters, and will always strive to provide the best fit by biasing the atmospheric model parameters. On the other hand, we found the L\,-\,retrieval to be highly sensitive to such effects and provide a significantly better safeguard against such systematic offsets.
    As more suitable instruments become available in the future, the field will move rapidly towards multi-instrument atmospheric retrievals. The L\,-\,retrieval approach described here may offer an optimal solution to interpret  multiple data-sets, taken at different times and/or with different instruments. 
    \newline     
    \newline
    \textbf{Software:} Iraclis \citep{Tsiaras2016}, TauREx3 \citep{al-refaie_taurex3}, PyLightcurve \citep{tsiaras_plc}, ExoTETHyS \citep{morello_exotethys}, Astropy \citep{astropy}, h5py \citep{hdf5_collette}, emcee \citep{emcee}, Matplotlib \citep{Hunter_matplotlib}, Multinest \citep{Feroz_multinest}, Pandas \citep{pandas_team}, Numpy \citep{oliphant_numpy}, SciPy \citep{scipy}.
    \newline     
    \newline
    {\large Acknowledgements}
    
    The authors would like to thank the significant input from the anonymous reviewer in improving the quality and presentation of the manuscript. 
    
    This project has received funding from the European Research Council (ERC) under the European Union’s Horizon 2020 research and innovation programme (grant agreement No 758892,ExoAI) and under the European Union's Seventh Framework Programme (FP7/2007-2013)/ ERC grant agreement numbers 617119 (ExoLights). Furthermore, we acknowledge funding by the Science and Technology Funding Council (STFC) grants: ST/K502406/1, ST/P000282/1, ST/P002153/1 and ST/S002634/1.
	{\small
		\bibliographystyle{aasjournal}
		\bibliography{main} 
	}
	

\end{document}